\documentclass[sigconf,anonymous=false]{acmart}
\usepackage{microtype}
\usepackage{url}
\newcommand{\code}[1]{{\footnotesize\url{#1}}}

\acmConference[VX-One '26]{Security \& ML Workshop (draft for submission)}{October 2026}{}
\acmYear{2026}
\renewcommand\footnotetextcopyrightpermission[1]{}

\begin{document}

\title{VX-One: Typed, Calibrated Decision Heads for Real-Time Perimeter Triage\\
  {\normalsize An empirical study of unknown-attack honesty, cascade gray zones, and calibration under drift}}

\author{Juan S. Santillana}
\email{juan@jsantillana.com}
\affiliation{\institution{Independent research}\city{}\country{}}

\begin{abstract}
Perimeter triage faces a latency/semantics trade-off: signature engines are fast and format-safe but blind to unseen attacks, while generative models are semantic but slow and produce unparseable output. In late 2026 a class of commercial \emph{System One} decision models (TypeSafe Jev; Microsoft-Decision-1) began selling typed decision primitives---binary gates, closed-set choices, ordinal scores---as product APIs, with no prior security-domain implementation or evaluation. We present \emph{VX-One}: a small multi-head network (100K--10M parameters) whose outputs \emph{are} these primitives—a calibrated \emph{gate}, a \emph{choice} over a closed candidate set with an explicit \texttt{unknown} option (a trained class where the benchmark provides an unknown population, abstention semantics elsewhere), and an ordinal \emph{score} (conditional ordinal regression), trained multi-task, temperature-calibrated per head ($\sim$0.7--1.7\,ms single-event latency p50 on CPU). We evaluate it on three public benchmarks (UNSW-NB15, CIC-IDS2017, EMBER2024) against calibrated XGBoost, Random-Forest, MLP, and Kitsune baselines, and measure a two-phase cascade (Suricata with 52{,}674 Emerging-Threats rules $\to$ VX-One) over five days of CIC-IDS2017. Our findings are deliberately double-sided: (i) on UNSW-NB15 in-distribution the gate \emph{ties} the best baseline (VX 0.9837$\pm$0.0002 vs MLP 0.9835$\pm$0.0013 across three seeds) and VX-One wins clearly under temporal drift on CIC-IDS2017 (0.734$\pm$0.022 vs.\ 0.639$\pm$0.027, paired $\Delta{=}{+}0.095$), but calibrated XGBoost wins multi-class \emph{attribution} on all three benchmarks; (ii) under leave-one-family-out the gate detects never-seen families on 9/9 UNSW folds yet drops below chance on CIC's \emph{recon} family---and on UNSW the unseen-family gate is a \emph{tie} with the strongest baselines (fold medians 0.987--0.991), while on stratified CIC the trees' gate transfers better and we report that as a loss, an abstention threshold calibrated on one benchmark transfers poorly to another; (iii) the signature tier's gray zone is real, not a reconstruction artifact: Suricata with the full ET Open ruleset resolves $\sim$0.1\% of daily attacks on flow reconstructions and only 0.2--2.2\% on the original payload-bearing PCAPs (0.45\% DDoS, 0.06\% PortScan on the wave day); a strictly by-day Phase 2 recovers only the DoS/DDoS share of that zone (0.34 recall, 0.97 precision on the wave day---the PortScan remainder is temporal aggregation, outside per-event scope), and a family-covered Phase 2 sets the $\ge$0.977 upper anchor; (iv) probability calibration collapses under temporal drift (ECE 0.04$\to$0.40) for \emph{every} method while ranking partially survives, and the drift is open-set (a family absent from training), which defeats label-shift estimators by assumption. We contribute not novel estimators but a measured account of where typed calibrated decisions help perimeter triage, and where they fail.
\end{abstract}



\maketitle

\section{Introduction}
\label{sec:intro}

Operating a mitigation policy on the network boundary requires, simultaneously: millisecond-scale decisions on the line of fire, calibrated confidences that a controller can threshold, and honest behavior on events that training has never seen. Three incumbent technology classes each miss at least one requirement.

\begin{enumerate}
  \item \textbf{Deterministic engines} (Snort~\cite{roesch1999snort}, Suricata~\cite{suricata}, Zeek/Bro~\cite{paxson1999bro}, YARA~\cite{yara}) are fast and format-safe but structurally blind to zero-day and low-and-slow offensives; the recurring failure mode is a rule explosion that no operations team can maintain.
  \item \textbf{Text LLMs for security} (e.g., SecureBERT~\cite{aghaei2022securebert})---~\emph{bidirectional encoders, not autoregressive models}, a conflation we correct---bring domain semantics, but as fine-grained decision engines on the line of fire they are heavy, and decoder-style LLM agents add generation latency and output-format fragility that a mitigation controller cannot tolerate.
  \item \textbf{Classical ML for intrusion detection} performs well on tabular attribution---XGBoost~\cite{chen2016xgboost} wins our attribution benchmarks---but treats every event as a membership test, producing confident misattributions on families the model never saw (Section~\ref{sec:e2}).
\end{enumerate}

During October 2026, a commercial class of models positioned exactly at this gap began to ship: \emph{System One} decision models, sold as APIs that take a structured \texttt{state} plus typed questions and return calibrated option probabilities (TypeSafe's Jev~\cite{jevdocs,typesafejev}; Microsoft-Decision-1~\cite{msdecision1,mslearn-decision1}). Independent academic evidence now covers JEV's language-evaluation routing~\cite{li2026jevjudge}; no security-domain assessment of the full triad existed.

This paper occupies the gap \emph{by building the primitives ourselves} rather than remoting into a vendor API. VX-One maps a fixed-size event state to three typed decisions in a single forward pass, then evaluates the whole pipeline---including a real signature engine in front of it---on public data with anti-strawman baselines. The contributions are:

\begin{itemize}
\item \textbf{A typed decision engine} (Section~\ref{sec:design}): shared encoder + three primitive heads---\emph{gate} (calibrated sigmoid), \emph{choice} (softmax over a declared closed set with an explicit \texttt{unknown} option: a trained class where the benchmark provides an unknown population, abstention semantics elsewhere), \emph{score} (CORN conditional ordinal regression~\cite{shi2023corn})---with per-head temperature calibration on validation data, trained multi-task with a homoscedastic weighting scheme~\cite{kendall2018multitask}.
\item \textbf{Unknown-attack honesty, measured} (Section~\ref{sec:e2}): leave-one-attack-family-out across 9 UNSW-NB15 and 7 CIC-IDS2017 families, plus 4 months of EMBER2024 family drift, reporting both unseen-family detection (gate AUROC) and \emph{attribution honesty} (how often the \texttt{unknown} option absorbs what the model should not claim to know).
\item \textbf{A measured gray zone} (Section~\ref{sec:e5}): Suricata 7 with the live Emerging-Threats ruleset resolves only $\sim$0.1\% of daily attacks over flow-reconstructed CIC-IDS2017 traffic and only 0.2--2.2\% over the original payload-bearing PCAPs---the gray zone survives full payload visibility. Under the deployment regime (a strictly by-day Phase 2), the cascade recovers only the in-scope DoS/DDoS share of the zone (0.34 recall, 0.97 precision on the wave day; the PortScan remainder is temporal aggregation, outside per-event scope by this paper's own account), and the $\ge$0.977 plateau is the family-covered (in-distribution) upper anchor---at the cost of a measured precision drop on web-attack days where Phase~1 is entirely blind.
\item \textbf{Negative results, measured} (Sections~\ref{sec:e1}--\ref{sec:e6}): calibration collapses under drift for every method (ECE 0.04$\to$0.40); class-weight tilts that help weak-label attribution can \emph{hurt} it on saturated benchmarks; a fixed abstention threshold is not portable across calibrations; dropping the \texttt{unknown} option raises macro-F1 by up to $+11.2$ points---the honesty/utility trade-off is real and quantifiable; and the wave-day drift is open-set (a family absent from training), which violates the assumption BBSE/EM need (P(x|y) invariance), so the new day's composition prior is not post-hoc-recoverable by label-shift machinery (E8).
\end{itemize}

Our heads are standard estimators; we say so. The claim is about their \emph{packaging and evaluation}: event-level honesty instead of confident misattribution, measured end-to-end on data with audited labels~\cite{engelen2021,lanvin2022errors}.

\section{Related Work}
\label{sec:related}

\paragraph{System One decision models.} Microsoft~\cite{msdecision1,mslearn-decision1} and TypeSafe's Jev~\cite{jevdocs,typesafejev} define and commercialize the primitive triad (binary gate / closed-set choice / ordinal score); their headline benchmarks are industry claims. Independently, Li et al.~\cite{li2026jevjudge} evaluate JEV's decision-only, confidence-routed language-evaluation cascade: its workload-specific threshold audit complements, but does not test, network telemetry. To our knowledge VX-One is the first academic security evaluation of the full triad; we define our own formal objects and use vendor names only for attribution.

\paragraph{Single-pass inference.} ``Non-autoregressive'' originates in neural machine translation~\cite{gu2018nat}: full output in one forward pass, no decoding loop. We use the term strictly for architecture (shared trunk, parallel typed heads). We do not claim novelty from single-pass-ness---BERT-family classifiers are also single-pass---and our baselines are strong calibrated tabular learners on the same feature stage; the differentiator is typed calibrated outputs, not one-pass-ness alone. Fast early-exit decision layers for per-event actions predate this design in spirit---the Viola--Jones cascade~\cite{violajones2001} is their ancestor.

\paragraph{Calibration and abstention.} Platt scaling~\cite{platt1999}, multiclass probability transforms~\cite{zadrozny2002}, empirical per-family calibration gaps~\cite{niculescu2005}, and temperature scaling~\cite{guo2017calibration} define our post-hoc protocol; conformal prediction~\cite{vovk2005,angelopoulos2021} is an optional wrapper we defer as a guarantee, but audit empirically on the temporal split (Appendix~\ref{app:formal}), noting that distribution-free coverage claims need exchangeability that temporal splits violate~\cite{barber2023}. Conformal risk control~\cite{angelopoulos2024crc} generalizes coverage to a bounded decision risk and is the right comparison class for our abstention policy, but it needs an exchangeable calibration stream and a monotone risk; the wave-day shift provides neither, so we measure risk--coverage empirically and defer the guarantee instead of claiming it. Chow's reject option~\cite{chow1970} and selective classification~\cite{geifman2017} ground our abstention semantics: with symmetric rejection cost the Bayes-optimal threshold is $1-\lambda_r = 0.5$, which is the threshold we initially set, and we measure how poorly it transfers where the model is overconfident (Section~\ref{sec:e2}).

\paragraph{Signature engines and their limits.} Snort~\cite{roesch1999snort}, Zeek~\cite{paxson1999bro}, YARA~\cite{yara}, and Suricata~\cite{suricata} anchor the deterministic tier. The DARPA-critique tradition~\cite{mchugh2000,mahoney2003} set the standard for dataset skepticism; audits of CIC-IDS2017~\cite{engelen2021,lanvin2022errors} document systematic labeling errors. We use both stratified and temporal splits and audit against the corrected labels.

\paragraph{ML on network traffic and malware.} Kitsune~\cite{mirsky2018kitsune} is the canonical packet-level unsupervised ensemble; we adapt it to flows (documented) and report its Platt-calibrated but weakly discriminative AUROC ($\approx$0.61) as a concrete instance of ``calibration creates no signal''. Boosted trees~\cite{chen2016xgboost} are the attribution baseline our engine \emph{loses to} on all three benchmarks---we calibrate them with the same protocol to make the comparison non-strawman. EMBER2018~\cite{anderson2018ember} and EMBER2024~\cite{joyce2025ember}, UNSW-NB15~\cite{moustafa2015unsw}, and CIC-IDS2017~\cite{sharafaldin2018cicids} supply the data tiers. Viola--Jones~\cite{violajones2001} is the ancestor of our second-phase-only-when-needed cascade.

\section{VX-One Design}
\label{sec:design}

\subsection{State object}
Each telemetry event is compiled to a flat state
\[
S \;=\; \bigl(x_{\mathrm{num}} \in \mathbb{R}^{d},\; x_{\mathrm{txt}}\bigr),
\]
where $d \in \{39 \text{ (UNSW)}, 77 \text{ (CIC-IDS2017)}, 2568 \text{ (EMBER2024)}\}$ and $x_{\mathrm{txt}}$ are protocol-level categorical fields (proto, service, state) when the benchmark provides them (UNSW only). The feature stage $\phi$ (median imputation of non-finite values with a row flag; signed $\log 1p$; robust scaling by train median/IQR; clip to $\pm8$) normalizes arbitrarily heavy-tailed benchmark features into a fixed-size vector. Everything is byte-level reproducible from frozen JSON configs with a single \texttt{SEED=42}.

\subsection{Trunk and typed heads}
A small pre-norm MLP trunk ($L\in\{2,3,4\}$ blocks of [Linear $\to$ LayerNorm $\to$ GELU $\to$ Dropout 0.1], residual from block 2) maps $\phi(S)$ plus text-lookup embeddings to a shared representation $h = f_\theta(S)$, feeding three typed heads---one per primitive:

\begin{align}
\text{\emph{gate}}: \quad & \operatorname{gate}(S) = \sigma(w_1^{\top} h + b) \in [0,1]; \tag{Gate}\\[-2pt]
\text{\emph{choice}}: \quad & \mathbf{p} = \mathrm{softmax}(W_C h), \quad \mathcal{C}_{\mathrm{eff}} = \mathcal{C} \cup \{\texttt{unknown}\}; \tag{Choice}\\[-2pt]
\text{\emph{score}}: \quad & \pi_k = \prod_{i\le k} \sigma_i(h), \tag{Score}
\end{align}
where the Score head is conditional ordinal regression (CORN~\cite{shi2023corn}) over four severity levels---benign, low, high, critical---and the choice head is a closed candidate set declared per dataset ($\mathcal{C}_{\mathrm{eff}}$ is the \emph{effective} set: where the training data offers a real \texttt{unknown} population, $\mathcal{C}$ already contains it as a trained class; where it does not, the union is realized at decision time as abstention, as described below). We deliberately use \emph{neutral} head names rather than vendor naming conventions. Trunk depth/width by scale ($L,w$): $2/224$ (Nano), $3/576$ (Small), $4/1280$ (Base), with reduced input projections on EMBER2024 (whose 2568-dim feature projection dominates the parameter budget); verified parameter counts on UNSW-NB15 are 97K / 914K / 5.9M, inside a hard $100\mathrm{K}/1\mathrm{M}/5$--$10\mathrm{M}$ budget that is enforced by unit test. The choice head's open-set semantics are \emph{hybrid}: where the training data offers a real \texttt{unknown} population (EMBER2024's 2.19M-file unknown ground truth), the option is a trained class; where it does not (UNSW, CIC), unknown is realized as abstention at decision time via $(\max_k p_k < \tau_{\mathrm{abs}})$, with $\tau_{\mathrm{abs}}$ and any composed policy belonging to the mitigation controller, not the network.

\subsection{Training formulation}
Multi-task loss
\[
\mathcal{L} ~ = ~ \sum_q \mathcal{L}_q,
\]
with a modified homoscedastic balancing~\cite{kendall2018multitask}: per-task reference uncertainty
with a modified per-task reference uncertainty so that task-scale differences (ln~$K$ vs.\ ln~2 for Choice vs.\ Gate) do not dominate the initialization. Per-class loss weights on Choice use sqrt-normalized, mean-1 tilts $\omega_k$ with an \emph{analytic de-tilt} in inference ($z \mapsto z - \log\omega$) applied before calibration, so the recovered posterior remains population-correct. The gate loss is \emph{unweighted} BCE: weighting shifts the optimal logit by $\log(\omega_+/\omega_-)$, an offset temperature calibration cannot undo. Training uses AdamW (weight decay $10^{-2}$ on 2D weight matrices only), linear warmup ($5\%$) + cosine decay, gradient clipping at 1.0, mixed precision, and early stopping (patience 4) on validation loss. All backbone hyperparameters are frozen across datasets and scales, with a single configuration table never tuned per dataset.

\subsection{Decision controller semantics}
\label{sec:gates}
The controller composes three binary gates from calibrated head outputs: the trained gate ($p^{\mathrm{cal}} \ge \tau_{\mathrm{det}}$, default 0.5); a drop gate on the exact CORN marginal $C_2 = \mathbb{P}(\mathrm{sev}\ge 2 \mid x)$; and a low-urgency gate on the \texttt{recon} coordinate of the choice head. The severity rubric makes both implications exact---$\mathrm{sev}\ge 2 \Rightarrow \text{attack}$ and $\text{recon} \Rightarrow \mathrm{sev}{=}1$---and they are verified with zero rule violations on 3.09M labeled flows, so each gate's target event coincides with exactly the event its own head estimates.

\paragraph{Gates are single-head functionals, by design.} For single-head functionals, post-processing (de-tilt, temperature, thresholding) keeps the calibration error within a bound determined by the head's own deficiency (Appendix~\ref{app:formal}, Lemma~1). Across heads, \emph{no useful composition bound exists}: two perfectly-calibrated heads can combine into a joint policy whose calibration error is $1/4$ in the worst case, whatever the marginal ECEs say (Appendix~\ref{app:formal}, Prop.~3)---and the $A{=}B$ case that attains it (two heads estimating the same underlying question) is the realistic one in security. The controller therefore composes \emph{boolean decisions} between gates and never \emph{probabilities} between heads; any future composed gate must be rubric-reducible to a single head, or measure its own joint calibration.

\paragraph{Thresholds are cost statements, not constants.} With uniform misclassification cost and escalation cost $\lambda_r$, the Bayes rule with rejection abstains iff $\max_k p_k < 1-\lambda_r$~\cite{chow1970,cortes2016rejection}: $\tau_{\mathrm{abs}}=0.5$ \emph{is} the statement $\lambda_r=0.5$ (abstention priced at half the cost of any wrong commitment), and $\tau_{\mathrm{det}}=0.5$ \emph{is} the statement that missing an attack costs the same as a false alarm. For operators who hold different cost ratios, the mapping is the standard cost-sensitive one, $\tau^{*} = c_{\mathrm{FP}}/(c_{\mathrm{FP}}{+}c_{\mathrm{FN}})$~\cite{elkan2001}---e.g., a 10:1 miss:alarm ratio implies $\tau^{*}\approx0.091$; we keep 0.5 as the declared default and expose the mapping rather than tune silently. General policies with a full cost matrix live in the controller---max-prob abstraction can deviate from true Bayes when costs are asymmetric (an event with $\max_k p_k$ above threshold may have lower expected cost elsewhere)---and we measure the deviation of the abstraction on validation data rather than assume it away.

\section{Experimental Setup}
\label{sec:setup}

\paragraph{Data.} \emph{UNSW-NB15}~\cite{moustafa2015unsw}: 149{,}039 / 26{,}302 / 82{,}332 flows (official split); 9 attack families. \emph{CIC-IDS2017}~\cite{sharafaldin2018cicids} (MLCVE-corrected variant): 2.83M flows; two splits, stratified (sanity) and by-day (temporal drift). \emph{EMBER2024}~\cite{joyce2025ember}: 3.85M binaries; temporal split by week, with the public challenge set held out; 2.19M ground-truth-unknown training rows. Label errors in the original CICIDS2017 are documented~\cite{engelen2021,lanvin2022errors}; on top of the audited variant, we repair raw-label encoding artifacts (U+FFFD mojibake in the web-attack labels) and a singular/plural naming drift (\texttt{Backdoor}), which a naive ingestion would silently route to the \texttt{unknown} class; micro-classes (11 Heartbleed, 36 Infiltration flows) are retained rather than dropped and stratified exactly. Every count is checked against the frozen schema in code before any experiment runs. Code, released results and weights: \code{https://github.com/vectrayx/vx-one} and \code{https://huggingface.co/jsantillana/vx-one}.

\paragraph{Baselines.} XGBoost (500 boosted trees, hist mode), Random Forest (300 trees, min-leaf 1), an MLP (2$\times$256), and Kitsune (9-autoencoder ensemble with feature bagging, Platt-calibrated on validation)---all trained with the \emph{same} feature stage, the same per-class loss weights where analogues exist, and the \emph{same} per-head temperature-scaling calibration as VX-One. This is anti-strawman discipline: boosted trees \emph{benefit} from the calibration fix~\cite{niculescu2005}, and we report them raw \emph{and} calibrated. A zero-shot API baseline from the commercial System One class is measured in Section~\ref{sec:e7}; LOAO controls for the same baselines and simple OOD detectors are added in Section~\ref{sec:e2}.

\paragraph{Metrics.} Detection: AUROC and \textbf{FPR@95\%TPR} (the FPR at the threshold capturing 95\% of true attacks---an operational line-of-fire number that accuracy-style metrics hide). Attribution: label-aware macro-F1 (GT-unknown rows excluded by label, as they have no known counterpart). Calibration: 15-bin ECE, MCE, Brier, NLL, and reliability diagrams, reported on raw and calibrated heads. Latency: single-thread, batch=1, 3{,}000 timed events, model-exact hardware recorded with each result.

\paragraph{Infrastructure.} Training on serverless GPUs (L4) with atomic checkpointing and tested resumption; evaluation computed \emph{locally} from dumped raw logits so that all metric code is inspectable. 60 unit tests including end-to-end smoke training ($\mathtt{pytest}$, 60/60 green).

\section{Results}
\label{sec:results}

\subsection{E1/E3 --- Detection, attribution, calibration}
\label{sec:e1}

Table~\ref{tab:e1} summarizes the two-tier comparison on test (calibrated temperatures). Four findings:

\begin{table}[t]
\caption{Test-set comparison (calibrated). VX: VX-One at three sizes. FPR@95 is FPR@95\%TPR. All rows now carry three seeds (42/43/44); the table shows seed 42 and the text reports seed means. Seed $\sigma$ of the gate AUROC is $\le$0.0013 on UNSW and EMBER, 0.017--0.027 on by-day CIC, and 0 on saturated CIC. With three seeds VX is 0.9837$\pm$0.0002 on UNSW vs MLP 0.9835$\pm$0.0013, XGBoost 0.9821$\pm$0.0003, RF 0.9793$\pm$0.0001: the in-distribution gate is a tie with MLP and a small but consistent edge over the trees. Row-paired bootstrap tests ($B{=}300$, same resampled rows for both models) are reported in the text. $^{\dagger}$ = under temporal drift (its by-day macro-F1 column is seed-unstable: VX spread 0.24--0.45 across three seeds, and XGBoost's 0.8336 is a single-seed value on the same unstable metric --- Section~\ref{sec:e1}); $^{b}$ = detection plateau of saturation (all models $\approx$ 1.0). Full per-dataset table, including ECE columns and Kitsune/RF/MLP rows, in the released results.}
\label{tab:e1}
\small
\begin{tabular}{llccc}
\toprule
Dataset & Model & AUROC & FPR@95 & macro-F1 \\
\midrule
UNSW-NB15 & VX Small & \textbf{0.9834} & \textbf{0.1211} & 0.4571 \\
          & XGBoost  & 0.9818 & 0.1369 & \textbf{0.4968} \\
          & RF       & 0.9792 & 0.1399 & 0.4378 \\
          & MLP      & 0.9821 & 0.1279 & 0.4445 \\
          & Kitsune  & 0.6086 & 0.5617 & --- \\
\midrule
CIC stratified$^{b}$ & VX Small & 0.9999 & 0.0003 & 0.8826 \\
                     & XGBoost & 1.0000 & 0.0002 & \textbf{0.9416} \\
                     & Kitsune & 0.6227 & 0.6910 & --- \\
\midrule
CIC by-day$^{\dagger}$ & VX Small & \textbf{0.7112} & \textbf{0.5731} & 0.4518 \\
                       & XGBoost & 0.6372 & 1.0000 & \textbf{0.8336} \\
                       & RF & 0.6308 & 1.0000 & 0.4207 \\
                       & MLP & 0.5959 & 1.0000 & 0.4579 \\
\midrule
EMBER2024 & VX Small & 0.9969 & 0.0119 & 0.7549 \\
          & XGBoost  & \textbf{0.9981} & \textbf{0.0053} & \textbf{0.8094} \\
          & MLP      & 0.9970 & 0.0098 & 0.7733 \\
\bottomrule
\end{tabular}
\end{table}

\textbf{(i) VX-One wins detection under drift; in-distribution it is a tie.} On temporal CIC, VX-One tops AUROC and FPR@95\%TPR: seed means 0.734$\pm$0.022 vs XGBoost 0.639$\pm$0.027 (paired per-seed $\Delta$ $+0.095$, range $+0.069$ to $+0.142$), and a row-paired bootstrap on seed 42 gives $\Delta$AUROC $+0.074$ $[+0.073,+0.075]$. On UNSW in-distribution the gate is a tie with the best baseline by the seed-level criterion: VX 0.9837$\pm$0.0002 vs MLP 0.9835$\pm$0.0013 across three seeds. The row-paired bootstrap at seed 42 does show a small advantage over MLP ($\Delta$AUROC $+0.0013$ $[+0.0010,+0.0016]$) and over the trees (XGBoost $+0.0016$, RF $+0.0043$); because that gap is of the same order as the retraining spread of MLP, we call it a tie rather than a win. On EMBER the gate is a small, consistent loss to XGBoost (0.9965$\pm$0.0005 vs 0.9981$\pm$0.0001, paired $\Delta$ $-0.0017$), which we also report rather than hide. On CIC by-day \emph{every} method's achievable FPR@95 is degraded, but the trees' degradation is total (FPR@95 = 1.0 --- no threshold reaches 95\% TPR at nontrivial FPR), while VX-One retains a partially usable operating point. Attribution (macro-F1) inverts this: calibrated XGBoost wins on UNSW, stratified CIC, and EMBER2024---by 4--6 points. \emph{We report the inversion rather than hide it}: our contribution does not rest on attribution wins, and we are explicit about the loss. The paired tests cover the other metrics as well, with the same resampled rows ($B{=}300$, seed 42): on UNSW the VX-minus-MLP differences are $\Delta$FPR@95 $-0.007$ $[-0.012,-0.001]$, $\Delta$macro-F1 $+0.012$ $[+0.003,+0.023]$, $\Delta$ECE15 $-0.008$ $[-0.010,-0.006]$, $\Delta$Brier $-0.015$ $[-0.016,-0.014]$; against XGBoost the attribution sign inverts ($\Delta$macro-F1 $-0.040$ $[-0.055,-0.023]$) while calibration and ranking favour VX. Under by-day CIC the ranking gap is large ($\Delta$AUROC $+0.074$ $[+0.073,+0.075]$, $\Delta$FPR@95 $-0.427$ $[-0.429,-0.425]$), and on saturated CIC every difference is at the fourth decimal---the benchmark has no headroom left. Intervals are percentile bootstrap intervals over test rows and are conditional on the trained model; they quantify row-sampling uncertainty, not retraining uncertainty, which is why the seed-level criterion above is the one we use for verdicts.

\textbf{(2) Saturation makes stratified CIC a sanity benchmark, not a selling point.} All methods land at $\sim$1.0 with ECE $\le$0.001. Headline numbers on saturated benchmark splits are best treated as pipeline-correctness checks.

\textbf{(iii) Seed variance on the four main-table runs.} To size the noise floor, UNSW-Small, CIC strat-Small, CIC by-day, and EMBER-temporal were re-run from scratch with seeds $\{42,43,44\}$ and evaluated with the same offline evaluator. The seed-42 rerun reproduces Table~\ref{tab:e1}'s official UNSW row to the fourth decimal (determinism of the pipeline, not luck). Across seeds, AUROC moves $\le$0.002 everywhere except by-day CIC ($0.734\pm0.022$), and macro-F1 carries a real spread exactly where the benchmark is least stable: stratified CIC $0.831\pm0.090$ (one seed dips to 0.73 while its AUROC and calibration stay saturated) and by-day CIC $0.344\pm0.104$ (spread $0.24$--$0.45$), versus UNSW $0.456\pm0.001$ and EMBER $0.746\pm0.014$ (all $\pm$ values here are sample SDs, $ddof{=}1$): \emph{detection} conclusions are seed-robust; \emph{attribution} conclusions on CIC are not, at any scale---the attribution claims lean on UNSW and EMBER.

\textbf{(iv) Kitsune: calibration does not equal signal.} With Platt calibration on validation, Kitsune is well-calibrated (ECE 0.037) yet near-chance on discrimination (AUROC 0.608). Its ensemble-of-autoencoders signal does not survive hostile-stage re-fit to benign-only data---we re-fit \emph{its} feature stage to its own training distribution (benign only), as a real unsupervised deployment would. Platt on a useless score calibrates a useless probability.

Calibration (E3): all supervised methods' in-distribution ECE stays $\le$ 0.066 on UNSW-NB15 (Kitsune, near-chance, is treated in finding (iv)) but drift to 0.39--0.45 on temporal CIC --- the pattern expected under distribution shift \cite{ovadia2019trust}. The policy-level implication is precise: \emph{`the deployed gate's calibration feedback stays calibrated'} is a claim about the training distribution; it silently degrades exactly when novel-attack pressure appears, which is the moment the claim was bought for. Section~\ref{sec:e6} quantifies its interplay with, not independence from, class-tilt choices. Classwise calibration of Choice (val) stays $\le$0.007 on all four main runs, and the \texttt{recon} coordinate that feeds $g_{\mathrm{recon}}$ is calibrated at 0.003 (UNSW, $n{=}$1{,}574) and 0.0001 (CIC strat)---max-prob calibration does not imply per-class calibration, so this is a property measured here, not assumed.

\subsection{E2 --- Unknown-attack honesty}
\label{sec:e2}

Leave-one-family-out (LOAO): train without family $F$, test against $F$'s held-out rows. Table~\ref{tab:e2} reports unseen-family \emph{detection} (gate AUROC / FPR@95 without any retraining), \emph{attribution honesty} (how much the choice head attributes to \texttt{unknown}), and a control (macro-F1 on \emph{known} families, which must not collapse).

\begin{table}[t]
\caption{LOAO---unseen-family detection (gate) and attribution honesty (choice abstention). The control---macro-F1 on \emph{known} families---stays at 0.38--0.56 across folds, i.e., the family that is excluded does not degrade attribution of the others. Baseline LOAO runs (calibrated XGBoost/RF/MLP plus logistic and Mahalanobis OOD detectors) are summarized in the text and released in full. UNSW Small.}
\label{tab:e2}
\small
\setlength{\tabcolsep}{4pt}
\begin{tabular}{lcccc}
\toprule
Excluded family & gate AUROC & FPR@95 & abst.\ (\%) & $n$ test \\
\midrule
analysis        & 0.9992 & 0.0002 & 92.32 & 677 \\
backdoor        & 0.9989 & 0.0028 & 63.46 & 583 \\
dos             & 0.9941 & 0.0202 & 2.76  & 4{,}089 \\
exploitation    & 0.9550 & 0.4177 & 8.19  & 11{,}132 \\
fuzzing         & 0.8519 & 0.3899 & 28.41 & 6{,}062 \\
generic         & 0.9957 & 0.0128 & 49.96 & 18{,}871 \\
recon           & 0.9523 & 0.1777 & 28.23 & 3{,}496 \\
shellcode       & 0.9688 & 0.1305 & 41.80 & 378 \\
worms           & 0.9871 & 0.0642 & 22.73 & 44 \\
\bottomrule
\end{tabular}
\end{table}

\textbf{UNSW: the gate generalizes.} Nine of nine never-seen families are detected with gate AUROC $\ge$ 0.85, most above 0.95, while macro-F1 on known families is preserved (0.38--0.56). The abstention rate on the excluded family ramps 3--92\%---feature-dependent, not tune-dependent, which is exactly the honesty modulation we want (an unseen-but-gatelike family still realizes $\texttt{unknown}$ 28--92\% of the time instead of being absorbed into \texttt{normal}).

\textbf{CIC: the failures are structural, not accidental.} On stratified CIC-Small LOAO (Table~\ref{tab:e2cic}), the gate remains strong where families have distinctive flow-level signatures (\texttt{bruteforce} 0.967, \texttt{webattack} 0.979, \texttt{exploitation} 0.970, \texttt{dos\_ddos} 0.852), but drops below chance on \texttt{recon} (0.426) and near chance on \texttt{infiltration} (0.533, $n=5$): in 77 flow-statistics features, port scans from a datacenter look exactly like a benign port-touching client.

\textbf{Baselines under LOAO (measured; full table with per-fold bootstrap CIs in the released results).} The same nine UNSW folds and seven CIC folds were re-run for the calibrated tabular baselines (identical splits, feature stage, and val temperatures) plus two simple OOD detectors trained on the known families only: a binary logistic regression and a one-class Mahalanobis (Ledoit--Wolf) on normal rows. The honest UNSW picture is that \emph{no method dominates the unseen-family gate}: VX tops 4/9 folds (analysis 0.999, backdoor 0.999, dos 0.994, generic 0.996), MLP 3 (recon 0.963, shellcode 0.974, worms 0.991), XGBoost and RF one each (exploitation 0.974, fuzzing 0.870); fold medians are MLP 0.991, VX 0.987, XGBoost 0.974, RF 0.962; the binary logistic reaches 0.78--0.96 and one-class Mahalanobis 0.65--0.87; and \emph{every} method degrades on \texttt{fuzzing} (0.65--0.87, VX 0.852)---an artifact family with no flow-level signature for anyone. Per-fold bootstrap 95\% CIs are usually narrow but do not settle all close orderings (e.g., VX \texttt{shellcode} [0.963,0.974], XGBoost \texttt{analysis} [0.971,0.983], and the 44-row \texttt{worms} fold is wider); we therefore use the paired fold tests rather than CI width as the verdict. A fold-paired Wilcoxon signed-rank test (Holm-corrected over the three calibrated baselines) confirms the UNSW verdict: VX vs MLP median $\Delta$ $+0.0005$, 5/9 folds, $p{=}0.73$; vs XGBoost median $+0.0007$, 7/9, $p{=}0.16$ (Holm 0.49)---a tie. On CIC it does not: excluding the two degenerate folds, XGBoost beats VX on 5/5 folds (median $\Delta$ $-0.123$; exact signed-rank $p{=}0.0625$, the smallest attainable with five folds), so we report a loss on that benchmark rather than average it into a tie. On stratified CIC the ordering flips on the weak folds: XGBoost's gate transfers better than VX's exactly where VX degrades (botnet 0.905 vs.\ 0.750; recon 0.830 vs.\ 0.426), the simple OOD detectors collapse (logreg 0.16--0.77, Mahalanobis 0.00--0.57 across folds), and the trees reproduce VX's saturation pathology---max-prob abstention at $\tau{=}0.5$ is $\approx$0 on all CIC folds, with unseen rows attributed to \texttt{normal}. On UNSW the trees' max-prob frontier is real but different in kind: RF abstains (median 0.80) at 0.14 median false attribution, while XGBoost/MLP keep attributing (median false attribution 0.53/0.61 at 0.40/0.34 abstention)---an untyped softmax abstains by low confidence, not by an explicit unknown option, so its frontier moves with miscalibration (Section~\ref{sec:e1}) instead of declaring the new class. We conclude the LOAO gate comparison is a \emph{tie in AUROC terms} whose tie-breaks are dataset- and fold-dependent; the typed design's measurable advantages under LOAO are the trained \texttt{unknown} option where a population exists, per-head calibration with post-processing guarantees (Section~\ref{sec:e1}), and CIC-recon honesty below---not raw AUROC superiority.

\textbf{Scope: what a single-flow triage \emph{cannot} see.} This blindness is structural, and stating its boundary is part of the contribution: VX-One classifies \emph{isolated} events (one flow's statistics, one binary's features). Reconnaissance is by definition a \emph{temporal-aggregation} phenomenon---low-and-slow scans are invisible to per-event statistics and become detectable only in detection-by-behavior over time windows (event provenance, targets-per-time histograms, scanning divisibility). The perimeter architecture therefore does not rely on the per-event head for recon: Phase~1 (Suricata portscan signatures, which partially cover it) and any windowed detector sit in front of, or beside, the per-event gate (Section~\ref{sec:e5}); when Phase~1 is payload-blind, the residual recon coverage is declared honestly rather than inflated.

\textbf{The abstention threshold measured at the default is non-portable}: on UNSW, $\tau_{\mathrm{abs}}=0.5$ abstains 3--92\% on excluded families; on CIC, where calibration is saturated (Table~\ref{tab:e1}, ECE $\le$0.001), the same fixed threshold is practically non-selective (mean per-fold abstention $\approx0$); benign absorption is the dominant, not exclusive, outcome. Under-saturated models are correctly selective, over-saturated ones are confidently wrong.

\begin{figure*}[t]
\centering
\includegraphics[width=0.9\textwidth]{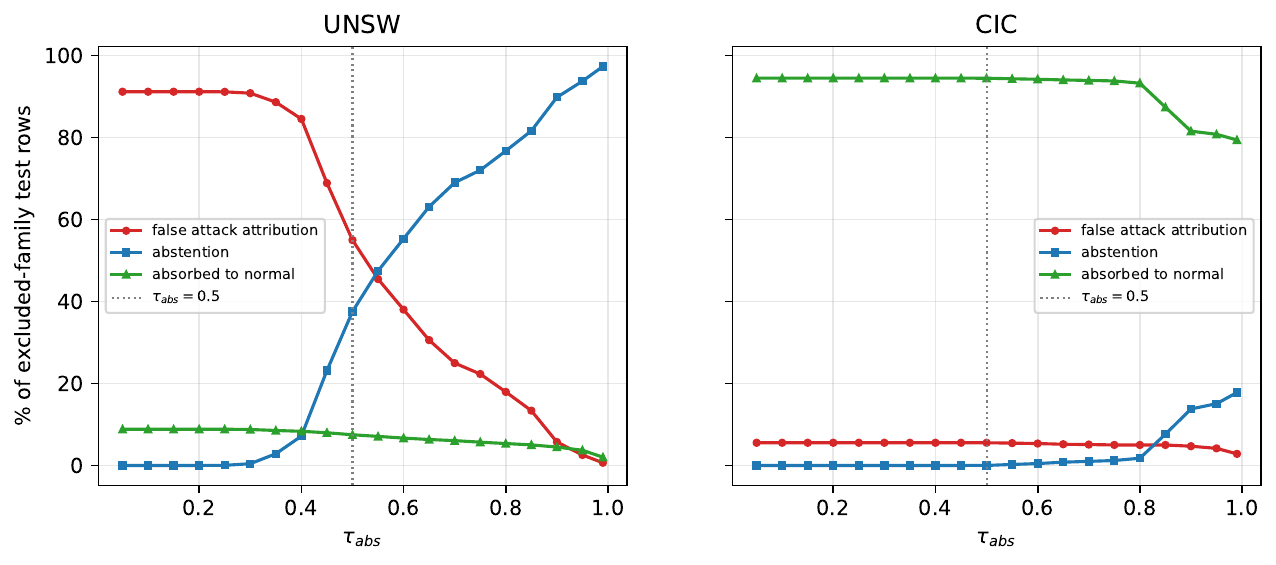}
\caption{Abstention sweep over $\tau_{\mathrm{abs}}\in\{0.05,\ldots,0.99\}$, mean over LOAO folds, on excluded-family test rows. \textbf{UNSW:} an operating frontier exists---abstention and false attack attribution cross near $\tau{=}0.5$, and large $\tau$ recovers honesty (97\% abstain, $<$1\% false attribution). \textbf{CIC:} no operating point---unseen-family events are absorbed into \texttt{normal} ($\sim$94\%) at saturated confidence (median max-prob $1.000$; median margin $1.000$), so no post-hoc threshold or margin policy can recover them.}
\Description{Two line panels of per-fold mean abstention rate and false-attack-attribution rate as the abstention threshold sweeps from 0.05 to 0.99. On UNSW the two curves cross near 0.5 and high thresholds yield 97 percent abstention at under 1 percent false attribution. On CIC the abstention curve stays near zero across the sweep and about 94 percent of unseen-family rows are attributed to the normal class at saturated confidence.}
\label{fig:tau}
\end{figure*}

\textbf{Sweep: a portable threshold does not exist (Figure~\ref{fig:tau}).} We swept $\tau_{\mathrm{abs}}\in\{0.05..0.99\}$ over the LOAO logits already computed (no retraining). On UNSW the sweep admits operating points: $\tau{=}0.5$ has a mean per-fold abstention rate of $37.5\%$ and mean per-fold false attack attribution of $55.0\%$, and rising to $\tau{=}0.99$ yields $97.3\%$ abstention at $0.66\%$ false attribution---a monotone honesty/utility frontier an operator can pick from. On CIC there is \emph{no} operating point: even at $\tau{=}0.99$ mean per-fold abstention reaches only $17.8\%$; the dominant failure mode is \emph{benign absorption} (mean per-fold $94.5\%$ attributed to \texttt{normal} at saturated confidence), rather than scattering across attack classes (mean $5.5\%$). Median max-probability on excluded rows is $\ge$0.999 on five of seven CIC folds (\texttt{recon}: $1.000$), and the margin alternative ($p_{(1)}-p_{(2)}$, median $1.000$ on \texttt{recon}) inherits the saturation: margin-based selective classification~\cite{geifman2017} fails for the same reason fixed thresholds do---post-hoc policies operate on the confidence distribution, and saturation has destroyed the information the policy needs. Rescuing a confidently-wrong model requires recalibration on in-scope data (per-window, online), not a sharper threshold; this is why abstention is a \emph{mitigation with preconditions}, not a guarantee. The constructive counterpart is measured (Appendix~\ref{app:formal}, E9b) under the same calibration contract: with delayed labels of the new family returned to training as an outlier pool ($\lambda{=}0.5$), re-training raises mean excluded-family abstention at $\tau{=}0.5$ from 0.375 to 0.879 on UNSW at a measured known-macro-F1 cost of $-0.007$; on saturated CIC the same recipe leaves the choice head's abstention at zero ($0.000\to0.000$; an earlier temperature-only readout suggesting otherwise was an estimator artifact we caught and correct) but recovers gate \emph{detection} (mean AUROC $0.783\to0.969$, \texttt{recon} $0.426\to0.991$)---honesty is teachable where the confidence distribution still carries information, and detection is teachable on the saturated regime; neither is a property a post-hoc threshold can replicate.

\begin{table}[t]
\caption{LOAO on stratified CIC (Small). Softmax confidence saturates: the fixed threshold is practically non-selective, and unseen families are predominantly attributed to \texttt{normal} with high confidence. $^{\dagger}$\,=\,degenerate folds ($n{=}1$ and $n{=}5$ test rows: the AUROC is a coin toss and is quoted for completeness only).}
\label{tab:e2cic}
\small
\begin{tabular}{lcccc}
\toprule
Excluded family & gate AUROC & FPR@95 & abst.\ (\%) & $n$ test \\
\midrule
botnet         & 0.7499 & 0.4038 & 0.00 & 295 \\
bruteforce     & 0.9674 & 0.0778 & 0.05 & 2{,}076 \\
dos\_ddos      & 0.8524 & 0.3394 & 0.00 & 57{,}103 \\
exploitation   & 0.9703 & 0.0297 & 0.00 & 1$^{\dagger}$ \\
infiltration   & 0.5329 & 0.8262 & 0.00 & 5$^{\dagger}$ \\
recon          & 0.4259 & 0.8685 & 0.00 & 23{,}840 \\
webattack      & 0.9792 & 0.1550 & 0.00 & 327 \\
\bottomrule
\end{tabular}
\end{table}

\textbf{EMBER2024 drift: trained-unknown as an honesty mechanism.} The EMBER run trains on weeks 0--45 and validates/tests on later weeks; family composition drifts measurably at test time. Gate-robustness is high (AUROC 0.997, FPR@95 0.012). The family-conditional table is the interesting one: \emph{berbew}, which grew 34.5K$\to$92K rows in test, retains recall 0.998; \emph{wacatac}, which mutated, falls to recall 0.129 with \textbf{85\%$\to$unknown}; \emph{grandoreiro} ($n=226$, almost absent from train) falls to 0.221 with 78\%$\to$unknown. The model does not reduce drift into false attribution; the \texttt{unknown} option absorbs what the model should not claim to know, and the gate keeps the operator's safety threshold honest on the detection side simultaneously. This is the strongest single argument for the trained-unknown semantics (and contrasts directly with Table~\ref{tab:e2cic} on two ways of switching unknown on/off in Section~\ref{sec:e6}).

\paragraph{What \texttt{unknown} means under LOAO.} The two unknown mechanisms carry the same typed meaning but are not the same estimator. On EMBER2024, \texttt{unknown} is a trained class with real ground truth (benign and tail malware), so its probability is an in-distribution rejection statement. Under LOAO on UNSW/CIC, the excluded family is not a candidate: abstention runs on max-softmax confidence over the \emph{seen} families, which is out-of-distribution similarity detection~\cite{hendrycks2017}---the classic max-softmax baseline, with its known overconfidence failure modes---not a calibrated probability about the excluded family. We therefore report unseen-family \emph{detection} (gate) and \emph{abstention} separately and never describe the latter as a probability of the unseen event. Both mechanisms lose informative honesty modes measurably (Tables~\ref{tab:e2}--\ref{tab:e2cic}); the design lesson is that distribution shift has no single calibrated fix---the honest object is the pair (gate, abstention), not one number.

\subsection{E4 --- Latency}
\label{sec:e4}

Measurements use single-events, batch=1, single-thread CPU, 3{,}000 timed events per model, on each model's exact released weights (Table~\ref{tab:e4}).

\begin{table}[t]
\caption{E4 latency (single-event, batch=1, 1 thread, CPU). Well within a $<$10\,ms design budget with margin $\ge 5\times$.}
\label{tab:e4}
\small
\begin{tabular}{lcccc}
\toprule
Scale & p50 ($\mu$s) & p95 ($\mu$s) & p99 ($\mu$s) & max ($\mu$s) \\
\midrule
Nano  & 675    & 715    & 773    & 853    \\
Small & 955    & 999    & 1{,}022  & 1{,}059  \\
Base  & 1{,}652  & 1{,}795  & 2{,}082  & 2{,}335  \\
\bottomrule
\end{tabular}
\end{table}

Interpretation: VX-One's typed decisions carry a p50 inference cost of $\sim$1\,ms on Small, making VX-One a \emph{per-event inline} decision component rather than a sampled-batch tool. The commercial System One class's headline latency claims are API-level; level-by-level comparison is not apples-to-apples, and we treat them as context, not as competition.

\subsection{E5 --- Two-phase cascade: how much of the line of fire is actually a gray zone}
\label{sec:e5}

We reconstruct five days of CIC-IDS2017 into PCAPs (documented reconstruction, no payload--- attacks' HTTP bodies are not recovered, so content-only signature rules are expected to underperform and this is the \emph{lower bound} they encounter on this traffic in the deployed position), run Suricata 7.0.10 with fresh Emerging-Threats Open rules (52{,}674 of them, concatenated) plus 8 of our own structural rules to level the playing field for header-only signals, then measure per family which attacks Phase~1 catches and what lands in the gray zone for Phase~2 to handle. Table~\ref{tab:e5} collates per day.

\begin{table*}[t]
\caption{E5 cascade over five days of CIC-IDS2017. ``PH1 acc'' is Phase 1 (Suricata) accuracy; ``PH2 acc'' is Phase 2 (VX-One) accuracy on that day's test rows; ``comb'' is the estimated combined cascade; P/R = precision/recall of the combined cascade. PH2/comb use the stratified-split (in-distribution) Phase 2; the by-day Phase 2 re-estimation on the wave day is in the text (P 0.971 / R 0.339) and Mon--Thu are not computable under it. \emph{Stats are estimates} with a documented independence assumption (Section~\ref{sec:limits}).}
\label{tab:e5}
\small
\setlength{\tabcolsep}{3.5pt}
\begin{tabular}{lrrrrrrr}
\toprule
Day & flows & attack & PH1 acc & PH2 acc & comb acc & comb P & comb R \\
\midrule
Mon 03 & 529{,}450 & 0        & 0.9989 & 0.9996 & 0.9985 & ---    & ---    \\
Tue 04 & 445{,}641 & 13{,}832 & 0.9681 & 0.9994 & 0.9985 & 0.958 & 0.996 \\
Wed 05 & 691{,}390 & 251{,}722 & 0.6359 & 0.9986 & 0.9979 & 0.997 & 0.997 \\
Thu 06 & 458{,}483 & 2{,}216  & 0.9942 & 0.9971 & 0.9961 & 0.552 & 0.9779 \\
Fri 07 & 702{,}713 & 288{,}785 & 0.5887 & 0.9979 & 0.9974 & 0.998 & 0.996 \\
\bottomrule
\end{tabular}
\end{table*}

\textbf{The headline number is the gray zone under payload opacity, not a universal estimate}. Suricata with ET rules over flow-reconstructed traffic detects only $\sim$0.1\% of daily attacks by volume; the \emph{gray zone} is 99.9\% of each day \emph{in this reconstruction}, and Phase 2 is what actually classifies it. \textbf{Lead with the deployment regime}: with the strictly by-day Phase 2 (the model an operator would actually have), the cascade recovers only the in-scope DoS/DDoS share of the wave-day gray zone---0.34 combined recall at 0.97 precision, arithmetically the 44\% family share times the 0.77 gray-zone recall---and the recovery value of Phase 2 concentrates in that single family, which Phase 1 itself barely touches (0.45\% of DDoS flows with full payload). The $\ge$0.977 recall / $\ge$0.95 precision plateau quoted next is the \emph{family-covered upper anchor}: it uses the stratified-split Phase 2, whose training contains rows of every attack family of every day (in-distribution per family)---the same saturated regime Section~\ref{sec:e2} criticizes. Re-estimating with the by-day Phase 2 (\texttt{results/e5/byday\_phase2}): the DoS/DDoS gray zone is still recovered (per-family recall 0.77), the PortScan share maps to a family index that is by construction absent from the by-day candidate set (temporal aggregation, outside per-event scope by this section's own account), and the combined estimate drops to recall 0.339 (precision 0.971); the Mon--Thu by-day legs are not computable with the current artifacts (those days are in the by-scheme training split), so Table~\ref{tab:e5} reports the in-distribution estimate for them. The payload-inclusive re-measurement above closes the remaining question---the gray zone persists even under a by-day Phase 2 with full payload visibility, because Phase 1's ruleset, not the reconstruction, is what leaves the zone. We are likewise explicit about what this experiment does and does not measure. It measures the Phase-2 recoverable share \emph{when Phase 1 is payload-blind}---the operational regime operators of encrypted traffic (TLS everywhere, no payload inspection in managed switches) know well---and \emph{not} the real-world gray zone of a deployment with full payload visibility, where Phase 1 resolves strictly more. Second, the cascaded numbers combine Phase 1 and Phase 2 events under a documented independence assumption at the shared-flow level; signature-triggered attacks also perturb flow statistics, so true overlap is nonzero and our combined recall is an estimate anchored at the payload-blind lower bound (Section~\ref{sec:limits}). \textbf{The Thursday failure is the real lesson}: on a web-attack day of 456K benign and only 2.2K attacks, Phase 1 is structurally blind without payload (precision 0.059), Phase 2 absorbs all of it, and the FP rate of Phase 2 across 456K benign rows dominates the combined precision (0.552). A cascade is not free: \emph{every} false flag from the blind-detection phase is multiplied by the innocent traffic around it; the correct operating posture there is Phase 2 as advisory (block with abstention, not auto-drop). This measurement---a live-tier engine's gray-zone share on a public benchmark under payload-only reconstruction---is, we believe, the crudest and most defensible novel number in this paper.

\textbf{The gray zone survives full payload visibility (measured on the original PCAPs).} To rule out that the gray zone above is an artifact of the payload-less reconstruction, we re-ran the same engine (Suricata 7.0.10, same ET Open daily ruleset, 52{,}706 rules, same eight structural rules) on the \emph{original payload-bearing CICIDS2017 PCAPs} and matched alerts to flows at the day-5-tuple level (the reconstruction's 2\,s/5\,s window protocol is not applicable here because the benchmark's recomputed flow timestamps drift $\pm$10\,min--$+$10\,h against the capture; reusing a flow's ephemeral port within the day biases benign-flagged upward, which we report as-is). With real payloads, ET Open resolves 2.2\% of Tuesday's brute-force flows, 2.2\% of Wednesday's DoS volume, 0.5\% of Thursday's web attacks, and on the wave day only 0.45\% of the DDoS and 0.06\% of the PortScan volume---while fully resolving the micro-families whose payloads are distinctive (Heartbleed 1/1, Infiltration 36/37). Between 2.2\% and 4.8\% of \emph{benign} flows alert on informational ET rules, so Phase 1's false-flag rate on real traffic is non-trivial. Three checks bound the matching's underestimation risk: every day-alert's 5-tuple is present among the day's enhanced flows (presence rate 1.000 on all five days, so no alert is lost to tuple absence), attack flows involving the documented attacker address (172.16.0.1) resolve at the same rate as attack flows overall (2.05\%/2.19\%/0.23\% on Tue/Wed/Fri), and the only declared bias (ephemeral-port reuse within the day) acts upward on benign-flagged, which we report as-is (\texttt{results/e5\_original/matching\_validation}). The gray-zone share of daily attack volume ($\ge$97.7\% on every day, and 99.5\% on the wave day) is therefore a property of the ruleset and the threat mix, not of the reconstruction: content signatures catch exploit payloads, not volume floods or scans.

\textbf{Risk--coverage of the cascade's own gating (measured).} We additionally measured the adjudication quality of the composed policy as a per-day risk--coverage curve (adjudicate ``attack'' iff Phase 1 flagged the flow or the Phase-2 gate exceeded $\tau$; abstain otherwise; risk = 1$-$precision of the adjudicated set; Phase-1 leg exact per flow, Phase-2 leg exact per test row, composed leg under the documented independence assumption of Table~\ref{tab:e5}). Findings: (i) the benign-only Monday is a null point --- \emph{every} adjudication is an error, so no operating point exists at risk $\le$30\%; (ii) on the Friday wave (DDoS+PortScan, 41\% of flows) the composed policy admits 0.416 of flows at risk $\le$5\%, the whole plateau supplied by Phase 2 (Phase 1 adjudicates 0.0004 of flows at precision 0.067: under payload opacity it contributes latency and provenance, not adjudication coverage, $\Delta\le$1 point); (iii) the small-attack Thursday (0.5\%) is where the cascade costs its own gate: Phase 1's 0.11\% flag rate at precision 0.059 corrupts every composed operating point, so nothing is admissible at risk $\le$10\%, and Phase 2's clean core covers only 0.44\% (min risk 0.3\%). This closes the next step of Section~\ref{sec:conclusion}.

\subsection{E6 --- Ablations}
\label{sec:e6}

Table~\ref{tab:e6} summarizes the ablation suite (full table, including the three-seed a7, and per-dataset effects in the released results). Five findings survive sanity checks and are worth their ink:

\begin{table}[t]
\caption{Ablations (ref = corresponding Small reference run).}
\label{tab:e6}
\small
\setlength{\tabcolsep}{4pt}
\begin{tabular}{llccc}
\toprule
Run & Effect tested & AUROC & m-F1 & ECE-ch \\
\midrule
UNSW ref    & ---             & 0.9834 & 0.4571 & 0.0496 \\
UNSW a1     & no text branch  & 0.9716 & 0.4079 & 0.0509 \\
UNSW a2     & single-heads    & 0.9784 & \textbf{0.4598} & 0.0552 \\
UNSW a3     & no calibration  & 0.9834 & 0.4571 & \textbf{0.0980} \\
UNSW a4     & no class tilt   & 0.9838 & 0.4545 & 0.0488 \\
UNSW a6     & no dropout      & 0.9839 & 0.4507 & 0.0451 \\
UNSW a7     & gate derived    & 0.9826 & \textbf{0.4727} & 0.0501 \\
\midrule
CIC ref     & ---             & 0.9999 & 0.8826 & 0.0003 \\
CIC a3      & no calibration  & 0.9999 & 0.8826 & 0.0007 \\
CIC a4      & no class tilt   & 0.9999 & \textbf{0.9313} & 0.0002 \\
CIC a5      & abstention off  & 0.9999 & 0.8826 & 0.0004 \\
CIC a6      & no dropout      & 0.9999 & 0.7360 & 0.0007 \\
CIC a7      & gate derived    & 0.9999 & 0.8616 & 0.0007 \\
\midrule
EMBER ref   & ---             & 0.9969 & 0.7549 & 0.0076 \\
EMBER a3    & no calibration  & 0.9969 & 0.7549 & 0.0112 \\
EMBER a4    & no class tilt   & 0.9969 & 0.7506 & 0.0061 \\
EMBER a5    & unknown off     & 0.9961 & \textbf{0.8674} & 0.0022 \\
EMBER a6    & no dropout      & 0.9966 & 0.7821 & 0.0081 \\
\bottomrule
\end{tabular}
\end{table}

\textbf{Text branch (a1).} On UNSW the protocol/service/state lookups are worth $\sim$5 macro-F1 points and nontrivial AUROC. UNSW's original feature set is partly categorical field values (proto, service, state), so the text branch supplies exactly the vocabulary a numeric encoder cannot infer. On EMBER2024 the ablation is an identity by construction: the 2{,}568 input features are already structural PE statistics (histograms, byte-entropy, imports), and no text field exists to ablate---which is also why the EMBER runs drop the lookup tables from the input projection entirely.

\textbf{Calibration changes probability, never ranking (a3).} Temperature scaling leaves argmax and AUROC untouched and improves ECE (e.g., 0.0496$\to$0.0980 when removed, on the same logits). This is the expected signature of a well-behaved calibration layer, and worth saying plainly because calibration \emph{is} our semantic differentiator.

\textbf{Class tilts are not free (a4).} On saturated CIC, removing the tilt $\omega$ \emph{improves} attribution by $+4.9$ macro-F1 at the reference seed---the de-tilt pushes minority classes \emph{down} in a regime where the weighted cross-entropy had already pushed them up. Three-seed re-runs put the a4-CIC effect at $\Delta=+4.9/+4.3/+3.9$ points across seeds ($\sigma{=}0.005$), so the direction is stable and the magnitude is seed-tight. On weakly-distributed EMBER the tilt matters less. There is no free class-tilt knob; its effect is deployment-regime-dependent, and our own loss design contains a knob whose default is wrong for saturated datasets. We say so and flag the design revision.

\textbf{Dropout is dataset-dependent and, on saturated CIC, seed-unstable (a6).} On UNSW, zero dropout is a stable no-effect ($0.4507/0.4522/0.4481$ macro-F1 across three seeds, $\sigma{=}0.002$; the single-head merge a2 is equally stable, $\sigma{=}0.003$). On CIC, zero dropout spans macro-F1 $0.736/0.663/0.912$ across seeds ($\sigma{=}0.128$) against a $0.883$ reference---the $-14.7$-point cost of the reference seed is within seed noise, and one seed even lands above it. On EMBER it \emph{gains} $+2.7$ (the majority class dominates and early stopping regulates). No universal recipe; report per dataset \emph{and per seed}.

\textbf{Typing the gate: a contract in-distribution, an open-set safeguard where absorption dominates (a7).} We retrained UNSW and CIC-Small with a single family softmax and no dedicated gate. A7 derives a raw marginal $1-p(\texttt{normal})$ after family de-tilt, clips it, and fits it an independent binary temperature; Choice retains its own temperature, so the two share family logits but not final calibrated probabilities (same trunk and budget). In-distribution the derived gate ties the dedicated one on UNSW ($0.9829\pm0.0014$ vs.\ $0.9837\pm0.0002$; $\Delta$FPR@95 $+0.008\pm0.008$; 1/3 paired seeds) and ties on saturated CIC; attribution moves $+0.006\pm0.011$ and $+0.067\pm0.100$ macro-F1 (2/3 each), within the paired-difference SD rather than evidence of a stable gain. Under LOAO the pattern splits: UNSW is a descriptive tie (median $\Delta$AUROC $-0.001$, 2/9, unadjusted Wilcoxon $p{=}0.30$), while CIC is mixed: \texttt{dos\_ddos} ($0.60$--$0.63$ across three a7 seeds vs.\ typed seed-42 $0.85$) and \texttt{botnet} ($0.57$--$0.59$ vs.\ typed seed-42 $0.75$) lose, but \texttt{exploitation} is seed-unstable ($0.42/0.94/0.87$ vs.\ typed seed-42 $0.97$; seven-fold $p{=}0.30$). Thus the dedicated gate has no in-distribution accuracy win or uniform open-set advantage; where benign absorption dominates it can improve detection, as a separately calibrated binary primitive.

\textbf{Unknown is a measurable trade-off (a5-wide).} Removing the trained-unknown option from EMBER's 11-way choice improves macro-F1 by $+11.2$ points (10-way is more balanced). ``Honesty'' is not free accuracy; making the \texttt{unknown} option available is a deliberate spend.

\subsection{E7 --- The commercial System One baseline, measured}
\label{sec:e7}

We ran the zero-shot API comparison this cycle against \emph{two} commercial \emph{System One} deployments---TypeSafe's Jev, SKU \texttt{jev-1.13-free} (the same model as the commercial SKU, served through a zero-metered access tier) and Microsoft's \emph{Decision-1} (Azure AI Foundry, deployment \texttt{decision-1}, model \texttt{Microsoft-Decision-1}, preview)---both exposing the same primitive names publicly (single decision model, documented gate/choice/score questions; one API request per flow containing the three primitives over the identical frozen feature state we feed our heads). The sample is the protocol-fixed stratified draw (seed 42) of the UNSW test set: 1{,}000 benign and 1{,}021 attack flows, attack quota proportional to family with a floor of 15. The vendor numbers are in Table~\ref{tab:e7}; the replication is the finding: \emph{both} vendors' gates sit at chance on this distribution (Jev AUROC 0.568, Decision-1 0.544), both choice heads are equally degenerate over our frozen candidate set (macro-F1 0.069 and 0.066 vs our 0.534), and both severity MAEs are $2$--$3\times$ ours (0.809 and 0.969 vs 0.333). Decision-1's gate is nominally better calibrated (ECE15 0.118 vs Jev's 0.265)---but with AUROC 0.54 there is no signal for calibration to expose, so the better ECE calibrates noise. Vendor latency: Jev 208\,ms p50 / 591\,ms p95; Decision-1 757\,ms p50 / 20.4\,s p95 (preview-tier cold starts; 1.24M input tokens for the run, price not published for the preview tier at run time; Jev's run cost $\approx$2.2M input tokens, US\$0.09 at the vendor's published rate). \textbf{Call variance (measured, both vendors)}: for Decision-1, $N{=}5$ repeated calls over a 2{,}041-row stratified draw (10{,}205 requests; the main E7 table uses the 2{,}021-row draw) give AUROC $0.540\pm0.002$, choice macro-F1 $0.066\pm0.001$, severity MAE $0.960\pm0.001$, and a per-row gate-probability $\sigma$ of 0.022; 13.1\% of rows flip their choice argmax between calls, and the metrics do not move. For Jev, $N{=}5$ over a 451-row stratified subsample (1{,}563 requests through the 9-node fleet, $\sim$3.4 calls per row after the free tier's per-IP budget) give AUROC $0.597\pm0.017$, choice macro-F1 $0.078\pm0.006$, severity MAE $0.924\pm0.025$, with a per-row gate $\sigma$ of 0.007 and 1.1\% argmax flips---Jev is near-deterministic per call, Decision-1 is noticeably stochastic, and \emph{neither escapes chance}. The verdicts are properties of the models on this distribution, not of single-call noise. For Jev we additionally ran the ablation adding an explicit \texttt{unknown} option to the vendor's choice question---over the \emph{full} 2{,}021-row sample so the comparison is paired (identical rows; the only difference is the extra option)---which leaves the attribution picture unchanged and pushes the gate further below chance (AUROC 0.476 vs.\ 0.568 without the option; macro-F1 0.066 vs.\ 0.069), and the vendor assigns \texttt{unknown} to only 0.35\% of rows (0.2\% of attacks, 0.5\% of benign). The escape hatch is available and unused; with the failure now replicated across two independent vendors, the gap is caused neither by the absence of an abstention option nor by prompt sensitivity nor by one vendor's implementation, but domain adaptation---the generic triad zero-shot does not transfer to network-flow distributions. The paper's positioning therefore still \emph{does not lean} on beating a commercial model; E7 closes the follow-up by measuring it, and is reported as an external stress test rather than a contribution.

\begin{table}[t]
\centering
\caption{E7: commercial System One baselines, zero-shot, same frozen state and rows as our heads (2{,}021 stratified UNSW test rows, seed 42). Two independent vendors of the primitive triad.}
\label{tab:e7}
\begin{tabular}{lrrr}
\toprule
Primitive / metric & Jev & MS-Dec.-1 & Ours \\
\midrule
Gate AUROC & 0.568 & 0.544 & 0.981 \\
Gate FPR@TPR95 & 0.932 & 0.872 & 0.157 \\
Gate ECE15 & 0.265 & 0.118 & 0.097 \\
Choice macro-F1 & 0.069 & 0.066 & 0.534 \\
Choice accuracy & --- & 0.323 & 0.711 \\
Severity MAE & 0.809 & 0.969 & 0.333 \\
\bottomrule
\end{tabular}
\end{table}

\subsection{Discussion}
Three threads bind the results together.

\textbf{(1) The honest triage division of labor.} Where the benchmark is saturable (stratified CIC), trees win attribution and a typed cascade adds nothing but a latency reduction; where the deployment regime is adversarial (temporal drift, attack families that did not exist in training), the gate/unknown semantics earn their keep in \emph{detection robustness and attribution honesty} even while losing the calibration war (E3). The contribution is the measured boundary of where typed decisions help, not a uniform ``we win everywhere''.

\textbf{(2) Honesty is a distinct axis from accuracy.} Section~\ref{sec:e2} and the EMBER drift table are the clearest pair of the paper: a benchmark-optimal system (no \texttt{unknown} option, accurate operating threshold) maximizes macro-F1 while \emph{falsely attributing} 85\% of a mutated family; the honesty-first system gives up 11 macro-F1 points on the choice metric and absorbs the mutation into an honest abstention. These are different products under drift, and the macro-F1 column alone cannot distinguish them.

\textbf{(3) Saturation is a deployment regime, not a benchmark property.} The same dataset (CIC-IDS2017) produces a saturated benchmark under stratified splitting and a hard unsaturated one under by-day splitting, in both ranking and calibration. Any paper reporting ``F1 0.94'' on CIC-IDS2017-stratified is measuring the model's ability to fit a monotone flow of re-used capture artifacts; papers should report the by-day numbers alongside (we do both).

\section{Limitations}
\label{sec:limits}

\begin{itemize}
\item \textbf{Severity scores are rubric-derived}, not human-labeled; the Score head's ground truth is a documented family$\to$level mapping, so cross-level claims are approximations above benign/low/high/critical --- a familiar evaluation pitfall for ML-in-security benchmarks \cite{arp2022donts}.

\item \textbf{Benchmark data, audited} (Section~\ref{sec:setup}): LOAO measures generalization \emph{across benchmark families}, not robustness against real-world novel attacks; an honest reading of E2 is ``the gate generalizes \emph{more} than the choice head, on these families''---not ``we detect zero-days'' (on the closed-world assumptions of IDS-data evaluation, see \cite{sommer2011contextual}).

\item \textbf{Single-seed runs outside the four main tables.} The four main-table runs (UNSW Small, CIC strat-Small, CIC by-day, EMBER temporal) were re-run with seeds $\{42,43,44\}$: the seed-42 reruns reproduce the official numbers to the fourth decimal (reproducibility confirmed), ranking metrics move $\le$0.027 points, and multi-class macro-F1 on CIC carries a real spread ($0.83\pm0.09$ stratified; $0.34\pm0.10$ by-day---sample SD, drift instability, not implementation nondeterminism). The ablation table is now three-seeded for a2/a4/a6 (a2 and a4 seed-stable, $\sigma\le$0.005; a6 on CIC seed-unstable, macro-F1 0.66--0.91, $\sigma{=}0.128$, Section~\ref{sec:e6}); the remaining single-seed tables are per-family LOAO baselines, E4 latency, and the E5 cascade; the remaining single-seed comparisons, including OE and per-family LOAO extensions, are descriptive; even large gaps can be fold- or seed-specific (Section~\ref{sec:e6}), so they are not generalized beyond their stated folds.

\item \textbf{Phase 1 is a reference implementation} (Suricata + ET Open + 8 structural rules), not our C++ engine; the claims of Section~\ref{sec:e5} are about the \emph{architecture}. The reconstruction leg means payload-content rules sit at a lower bound by construction, but the original-PCAP leg (measured, Section~\ref{sec:e5}) shows the gray zone persists at full payload visibility (0.2--2.2\% daily resolution), so the architecture-level conclusion does not rest on the reconstruction. The ``combined cascade'' result is a documented two-sample estimate under an independence assumption that is exact for full-family-invisibility cases and optimistic where Phase 1 fails on exactly the stealthiest subset; we state it as an estimate, and the per-day F1/gray-zone table it rests on is a real measurement---whose Phase-2 leg uses the stratified-split (family-covered) Phase 2: re-estimated with the strictly by-day Phase 2, the wave-day combined recall drops from 0.996 to 0.339 (precision 0.971), so the size of the ``recovered'' gray zone, not only of Phase 1, depends on the split regime and is anchored at the family-covered upper bound.

\item \textbf{Calibration under drift} (ECE 0.40 on temporal CIC) is measured and now also audited by a walk-forward per-window recalibration experiment (E8, Appendix~\ref{app:formal}): with one-day-delayed labels, gate recalibration (temperature or 2-parameter Platt) recovers part of the loss (Brier 0.539$\to$0.504 on the hard day), but the Phase-2 non-parametric check sharpens the bound: deployable isotonic maps fitted on past windows transfer 4.0 pts of ECE inside the regime and nothing across it (0.5531$\to$0.5486), and even the rank-respecting day-self oracle only reaches a 0.260 \emph{hard-prediction} error (from the flat map's 0.445; metric correction in Appendix~\ref{app:formal})---the deployable ceiling is \emph{discrimination, not calibration}, and explicit label-shift correction does not change it: the wave-day shift is open-set (the PortScan family is absent from training), so BBSE/EM's consistency assumption $P(x\,|\,y)$-invariance is violated, and measured estimators recover $\widehat{q}\le0.04$ vs.\ true $0.555$ even in-scope (Appendix~\ref{app:formal}). Per-window recalibration is therefore a cheap partial fix. Its prescribed alternative is also measured (E9, Appendix~\ref{app:formal}): one to two passes of per-window continue-training with delayed day-1 labels recover day-2 gate AUROC to 0.80--0.84 (four passes regress to 0.76) and, with the recovered ranking, the day-self oracle's hard-prediction gap drops from 0.445 ($=1{-}q$, the flat-map floor) to 0.10--0.12, the ranking-implied hard error at AUROC 0.84---but as-run day-2 ECE stays at 0.554 and no delayed-label fit recovers the new day's composition prior, which lands on the abstention/coverage semantics of Section~\ref{sec:e2} rather than on any post-hoc map.

\item \textbf{Conformal wrappers are scoped, not promised, and now audited.} Exact distribution-free coverage applies only under exchangeability: on the official UNSW split and stratified CIC, split-conformal coverage is exact~\cite{papadopoulos2002,tibshirani2019}; on every temporal split, family-composition drift is real and measured, and coverage degrades at a rate bounded by the calibration--test total-variation gap~\cite{barber2023}. A domain classifier certifies that gap only \emph{from below} ($d_{\mathrm{TV}} \ge 1-2\,\mathrm{BER}$; Appendix~\ref{app:formal})---we therefore report empirical weekly coverage with exact binomial intervals and weighted variants~\cite{tibshirani2019}, and make no distribution-free coverage promise; the measured EMBER audit is in Appendix~\ref{app:formal} (binary set-coverage within $[0.895,0.935]$ across all twelve test weeks; certified drift $d_{\mathrm{TV}} \ge 0.15$--$0.24$).

\item \textbf{Cost-per-million} is not yet measured from real billing data; latency, not cost, is the number we claim.

\item \textbf{Text-field attack surface} (indirect prompt injection through \texttt{user\_agent}, URIs) is real~\cite{greshake2023injection}; V0.1 has no sanitization layer and our text-branch vocabulary lookup (bounded 135-slot vocabularies, not free max-k) is only partial mitigation.

\item \textbf{Vendor-API baseline is a two-vendor replication on one distribution} (Section~\ref{sec:e7}); we measured two commercial System One deployments (TypeSafe Jev, Microsoft Decision-1) zero-shot on one distribution (dense UNSW flows), not an exhaustive comparison across commercial decision models, prompt variants, or distributions---and the two vendors' agreement (both gates at chance, both choice heads degenerate) is the measured evidence, but the sample is still one domain. The honest reading is ``the generic triad zero-shot does not transfer to this domain without fine-tuning'', not ``the vendors cannot be fine-tuned to this domain''.
\end{itemize}

\section{Conclusion}
\label{sec:conclusion}

We built, trained, calibrated, and stress-tested a typed decision engine for perimeter triage---to our knowledge the first public-data security-domain implementation and evaluation of the \emph{System One} primitive triad---and report the boundary of its advantage rather than a uniform win: detection robustness and attribution honesty under unknown families and drift, a measurable gray zone of a live signature engine, and five cleanly-measured negative results (calibration collapse under drift, non-portable abstention thresholds, regime-dependent class tilts, unknown-option accuracy cost, and label-shift estimators failing by assumption violation under open-set drift). The following former next steps are already measured: the vendor-API comparison (Section~\ref{sec:e7} --- the commercial primitive triad zero-shot sits at gate-chance on network flows \emph{on both measured vendors} (Jev and Microsoft Decision-1), and an explicit abstention option is not enough to close the domain gap), the abstention-threshold sweep (Section~\ref{sec:e2}, Figure~\ref{fig:tau}---on saturated deployments no abstention threshold recovers honesty, so selective behavior must be bought with data and recalibration, not with post-processing), the 3-seed re-runs of the four main tables (seed-variance on ranking metrics $\le$0.002 everywhere except by-day CIC, and $\le$0.027 including it; Section~\ref{sec:limits}) and of the ablations (a2 and a4 seed-stable, $\sigma\le$0.005; a6 on CIC seed-unstable, macro-F1 0.66--0.91, $\sigma{=}0.128$---the reference-seed $-14.7$-point dropout cost is within seed noise; Section~\ref{sec:e6}), the walk-forward per-window recalibration under drift (E8, Appendix~\ref{app:formal}---a cheap partial fix whose Phase-2 non-parametric oracle check confirms the hard-day deployable ceiling is discrimination, not the calibration family: rank-respecting isotonic maps fitted on past windows recover 4.0 pts of ECE inside the regime and none across the shift, and the shift is open-set, so label-shift estimators are out of scope by assumption), its prescribed alternative, per-window re-training (E9, Appendix~\ref{app:formal}---one to two passes with delayed day-1 labels recover day-2 gate AUROC 0.56$\to$0.84 and re-enable oracle recalibration, while four passes regress and as-run ECE is unmoved: weight freshness buys discrimination, not calibration), and the risk--coverage curves of the cascade's own gating (Section~\ref{sec:e5}---on payload-blind traffic the coverage at fixed risk is supplied entirely by Phase 2, the benign-only day is a null point where no operating point exists, and the small-attack day is where Phase-1 false flags cost the gate itself), and the payload-inclusive Phase-1 re-measurement on the original PCAPs (Section~\ref{sec:e5}---the gray zone persists at full payload visibility: 0.2--2.2\% of daily attack volume resolved, 0.45\% DDoS / 0.06\% PortScan on the wave day, so the cascade architecture's premise survives the strongest possible Phase-1 engine under this ruleset). The remaining next steps are extending outlier exposure to cross-domain outlier pools (the delayed-labels pool, the $\lambda$ sensitivity, and the synthetic-noise ablation are measured in Appendix~\ref{app:formal}) and the fleet-based $N{=}5$ pass over Jev's \emph{full} sample (the 451-row subsample variance is measured, Section~\ref{sec:e7}). All code, configs, tests, and measured results accompany the paper.

\appendix
\section{Formal statements}
\label{app:formal}

Throughout, $\mathrm{cal}_A(f) = \mathbb{E}\big[\,|\,\mathbb{P}(A \mid f(X)) - f(X)\,|\,\big]$ is the $\ell_1$ calibration error of an estimator $f\in[0,1]$ for event $A$; the binned ECE (15 bins) is its empirical estimator~\cite{kumar2019}, with its known estimator caveats, which is why we co-report MCE and reliability diagrams and frame multiclass notions through the vector/classwise/confidence hierarchy~\cite{vaicenavicius2019}.

\paragraph{Statistical protocol (retrospective, fully specified).} Two uncertainty sources answer different questions. \emph{Row sampling} is summarized by a paired bootstrap ($B{=}300$ resamples of identical test rows); its percentile 95\% intervals are conditional on the trained model and descriptive, not verdicts. \emph{Retraining} is assessed with seeds 42/43/44 where both systems were rerun; a main-table comparison is a \emph{win} only when paired seed differences have one sign and exceed their spread, otherwise a \emph{tie}. Baseline LOAO tests use the nine UNSW folds and the five non-degenerate CIC folds, with Holm correction over the three calibrated baselines. The later a7 LOAO and OE extensions report all named folds (including small-$n$ folds explicitly) and their unadjusted fold-level Wilcoxon $p$-values as descriptive evidence, not family-wise claims. Threshold $\tau{=}0.5$, bootstrap size, folds, and seeds are fixed in the released scripts; this protocol was finalized during revision rather than preregistered.

\paragraph{Proposition 5 (label shift is a vector, not a scalar).} Let source and target differ only by label shift, $P_t(x\mid y)=P_s(x\mid y)$ and $q_y \neq p_y$ for some $y$. Then the calibrated target posterior is
$p_t(y\mid x) = p_s(y\mid x)\,\frac{q_y}{p_y}\big/\sum_{y'} p_s(y'\mid x)\,\frac{q_{y'}}{p_{y'}}$,
so a single temperature $T$ fitted on the source cannot in general calibrate the target: the correction is a per-class multiplicative tilt, and $T$ can represent it only when all ratios $q_y/p_y$ are equal (no shift). De-tilting by the \emph{target} prior is exact under the stated invariance; estimating that prior from unlabeled target data is where BBSE/EM enter, and their consistency fails as soon as a class absent from the source appears, because $P(x\mid y)$ is then undefined for that mass. \emph{Proof.} Bayes' rule gives $p_t(y\mid x) \propto P_t(x\mid y)q_y = P_s(x\mid y)p_y\,(q_y/p_y) \propto p_s(y\mid x)(q_y/p_y)$; the normalizer follows. A scalar $T$ rescales logits uniformly and cannot reproduce a vector tilt unless all ratios coincide. \hfill$\square$

\paragraph{Lemma 1 (closure under single-head post-processing).} If head $f$ is calibrated for an event $A$ of its own output and $h:[0,1]\to[0,1]$ is measurable, then
$\mathrm{cal}_A(h\!\circ\!f) \le \mathrm{cal}_A(f) + \mathbb{E}\,|h(f(X)) - f(X)|.$
\emph{Proof.} Let $m(v) = \mathbb{P}(A \mid f(X){=}v)$. Tower property: $\mathbb{E}[\mathbf{1}_A \mid h\circ f] = \mathbb{E}[m(f)\mid h\circ f]$. Jensen: $\big|\mathbb{E}[m(f)\mid h\circ f] - h(f)\big| \le \mathbb{E}\big[\,|m(f)-h(f)|\,\big|\,h\circ f\big]$. Triangle: $|m(f)-h(f)| \le |m(f)-f| + |f-h(f)|$; take expectations. \hfill$\square$

So de-tilt, temperature scaling, and thresholding do not destabilize a gate's calibration beyond the size of the transformation; the practical consequence is that each derived gate's calibration error is \emph{bounded by its own head's measured deficiency}, and E3 measures each gate's ECE directly anyway.

\paragraph{Lemma 2 (CORN chain rule: exactness and degradation).} With true conditionals $\pi_1(x) = \mathbb{P}(\mathrm{sev}\ge1 \mid x)$ and $\pi_2(x) = \mathbb{P}(\mathrm{sev}\ge2 \mid \mathrm{sev}\ge1,\,x)$: (i) $\mathbb{P}(\mathrm{sev}\ge2 \mid x) = \pi_1(x)\,\pi_2(x)$ \emph{exactly}---CORN conditions the second step on $\mathrm{sev}\ge1$, so there is no hidden independence assumption; (ii) for estimates $(q_1, q_2)$,
$\mathrm{cal}_{\mathrm{sev}\ge2}(q_1 q_2) \;\le\; \mathbb{E}\,|\pi_1 - q_1| + \mathbb{E}\,|\pi_2 - q_2|.$
\emph{Proof of (ii).} $\pi_1\pi_2 - q_1 q_2 = \pi_2(\pi_1 - q_1) + q_1(\pi_2 - q_2)$; triangle inequality with $0 \le \pi_2, q_1 \le 1$. The bound is on \emph{conditional deviations}, not marginal ECEs (conditioning on extra information changes the base rate~\cite{kleinberg2017}); hence the drop gate's ECE is measured directly rather than inferred from its conditional heads. \hfill$\square$

\paragraph{Measured (drop gate, this deployment).} The links' conditional ECEs do not determine the product's: UNSW-Small val gives ECE($q_1$){=}0.014, ECE($q_2\,\vert\,$sev$\ge$1){=}0.038, direct ECE($C_2$){=}0.026, and temporal CIC at test gives ECE($q_1$){=}0.42 and ECE($q_2$){=}0.14 with direct ECE($C_2$){=}0.11. Inferring the composed gate from its links mis-estimates it in either direction; the controller measures $C_2$ directly.
\paragraph{Proposition 3 (no-go: marginal ECEs do not compose).} Let $X$ be trivial and $f_1 \equiv a$, $f_2 \equiv b$ be perfectly calibrated for events $A, B$ with $\mathbb{P}(A){=}a$, $\mathbb{P}(B){=}b$ ($\mathrm{cal} = 0$ for both). The naive AND policy $g = f_1 f_2 = ab$ estimates $A\wedge B$ with
$\sup\,\big|\,\mathbb{P}(A\wedge B) - ab\,\big| = 1/4,$
attained at $a=b=\tfrac12$ with $A{=}B$ or $A\cap B{=}\emptyset$ (the joint law is free within the Fr\'echet bounds), and identically for the OR variant. Therefore no non-trivial function of the marginal ECEs bounds composed-policy calibration error; the attainable case $A{=}B$---two heads estimating the same question, e.g., ``is this an attack?''---is realistic in security, not pathological. This is why Section~\ref{sec:gates} restricts the controller to composing boolean decisions and forbids cross-head probability composition: every gate of v0.1 is a functional of exactly one head, verified rubric-reducible. \hfill$\square$

\paragraph{Lemma 4 (BER--TV; drift certification is one-sided).} For any two distributions $P, Q$, $d_{\mathrm{TV}}(P,Q) = 1 - 2\,\mathrm{BER}^{*}$, where $\mathrm{BER}^{*} = \inf_h \tfrac12\big(\mathbb{P}_P(h{=}1) + \mathbb{P}_Q(h{=}0)\big)$ is the optimal balanced error (the likelihood-ratio test attains error $(1-d_{\mathrm{TV}})/2$). Consequently a domain classifier with balanced error $\mathrm{BER}$ \emph{certifies} $d_{\mathrm{TV}} \ge 1 - 2\,\mathrm{BER}$: detecting drift quantifies it and plugs into the coverage-gap bound of~\cite{barber2023}; failure to separate certifies nothing---absence of evidence of drift is not evidence of exchangeability. \hfill$\square$

\paragraph{Conformal audit on the temporal split (measured, binary head).} Split-conformal on the EMBER gate's temperature-calibrated logit ($\alpha{=}0.1$; calibration on validation weeks 46--51, $n{=}120$k) reproduces the target coverage exactly at calibration (0.9000). Across the twelve test weeks: unweighted coverage $0.895$--$0.935$ (exact binomial intervals), benign-side $\ge$0.905 everywhere, malicious-side down to $0.884$ (week 60) --- the loss concentrates on the positive class. Per-week feature drift is certified from below: $d_{\mathrm{TV}} \ge 0.15$--$0.24$ (held-out BER; domain-classifier AUC 0.61--0.64, so the feature-space drift is real but modest). The weighted variant (clipped odds $[0.1,10]$, Tibshirani-style quantile) moves coverage by $\le 0.022$ with no consistent direction --- with estimated weights it is mitigation, not guarantee, and here it is unnecessary: what drifts over the four months is the family mixture (label shift, absorbed by \texttt{unknown}; Section~\ref{sec:e2}), not the benign/malicious feature distribution. The challenge set is excluded from these statements and reported separately as measured stress: its coverage on positives is 0.27 (by construction 98.7\% \texttt{unknown}-family, minimally evasive), with a median gate likelihood of 0.65 --- the detector keeps voting attack on most rows ($\ge$0.5 for 58\%) but without claiming the set-coverage it does not have.

\paragraph{Bayes rule with rejection (abstention thresholds).} With uniform misclassification cost and escalation cost $\lambda_r$, the expected cost of an action is $C(a\mid x) = 1 - p_a(x)$ under typed outputs, so Bayes abstains iff $\min_a C(a\mid x) > \lambda_r$, i.e., iff $\max_k p_k(x) < 1-\lambda_r$---Chow's 1970 rule as the 0--1-cost case~\cite{chow1970, cortes2016rejection}, with risk--coverage curves as the primary object~\cite{elyaniv2010, geifman2017}. Asymmetric matrices of cost live in the controller: max-prob abstraction can deviate from full Bayes there (a one-line counterexample: $p = (0.55, 0.40, 0.05)$ with a 3-fold penalty on confusing the second class yields abstention under full Bayes but attribution under max-prob), which is why the deviation is measured on validation data in the controller's calibration report rather than assumed away. Measured: with uniform costs the abstraction \emph{is} the Bayes rule (zero deviating rows in all three runs with a \texttt{normal} class); under an $m$-fold penalty on missed attacks, deviations appear in one direction only---full Bayes escalates where max-prob attributes---at 13\% of validation rows (UNSW, $m{=}3$) and up to 29\% among attacks (by-day run, $m{=}10$, $\lambda_r{=}0.1$); symmetric or moderate costs justify the default $\tau{=}0.5$, but ratio must re-derive the policy rather than inherit it.

\paragraph{Walk-forward per-window recalibration under drift (E8, measured).} On the CIC by-day run (test in capture order: Friday-DDoS $n{=}138$k, then Friday-PortScan $n{=}286$k), we re-fit the gate's calibration on the previous window only (day 1 falls back to the static val; day 2 uses day-1 labels, the honest delayed-labels regime), either as a temperature (1 parameter) or as a 2-parameter Platt map $p{=}\sigma(a\,\mathrm{logit}{+}b)$, and re-derive the FPR@TPR95 threshold from the same window's positives. On the DDoS day (AUROC 0.99) the oracle bound recovers almost all calibration error (ECE 0.127$\to$0.029, gate fitted on the day's own labels). On PortScan (AUROC 0.56) the same post-hoc moves are marginal: ECE 0.553 (static) $\to$ 0.526 (Platt with one day of fresh labels). A Phase-2 extension replaces the parametric family with non-parametric maps (artifacts in the repo, \texttt{results/exp\_f2/}): (i) \emph{deployable} isotonic regression fitted on val transfers on the within-regime day (ECE 0.127$\to$0.087, Brier 0.125$\to$0.056) but does \emph{not} transfer across the shift (day-2 ECE 0.5531$\to$0.5486; fitted on day-1 delayed labels 0.5521)---the failure is shift-specific, not a valley of the parametric family; (ii) the \emph{parametric} day-self oracle bound quoted as ECE 0.445 (temperature family) is family-limited: a day-self isotonic map reaches 0.260 (Brier 0.539$\to$0.178)---and both figures need a metric correction that we own, because it changes their meaning. The recalibration tables' gate ``ECE15'' is computed as the \emph{hard-prediction gap}: within confidence bins, $|\,\operatorname{mean}(\mathbb{1}[p\ge\tfrac12])-\operatorname{mean}(y)\,|$, the FP-vs-FN imbalance of the binarized map---not the standard top-label ECE of Section~E3 ($|\mathrm{acc}-\mathrm{conf}|$). Under that metric, \emph{any} map whose outputs all sit at or above 0.5 scores exactly $1-q_2=0.445$ on this day: the day-self temperature optimum (slope $a{=}0.06$, outputs $\sigma(0.06\,\mathrm{logit}+0.98)\in[0.51,0.80]$) and the constant-at-prior map both land on 0.4452. The ``0.445 plateau'' is therefore $1-q$---the floor of an all-1 predictor---not a calibration ceiling. Under the standard soft/top-label ECE the same maps are (near-)calibrated: the constant at the true prior has ECE 0.000 and Brier $q(1{-}q)=0.247$, and the day-self isotonic reaches ECE 0.000 with Brier 0.178. What the weak ranking (AUROC 0.56) actually bounds is the \emph{hard error}: the best monotone map binarizes to 26\% error (the tables' 0.260 is this quantity), versus 44.5\% for any flat map---no calibration can push hard error below the ranking-implied Bayes error, and flattening buys soft calibration only by surrendering all resolution (Brier $\to q(1{-}q)$). The deployability failure stands separately: nothing fitted on past windows transfers across the shift (soft ECE 0.5531$\to$0.5486; hard gap 0.553). The label-shift estimators, however, must be read with their scope: BBSE and EM/Saerens are consistent \emph{under label shift}, i.e., $P(x\,|\,y)$ invariant. The wave-day shift is open-set---the PortScan family is absent from the by-day training set, so $P(x\,|\,y)$ \emph{changes} and a new class appears as a point mass no posterior-based estimator can price without fresh labels. Their measured failure ($\widehat{q}_2 \le 0.04$ against a true prior of $0.555$, all estimators, day-1 delayed labels; \texttt{scripts/e8\_labelshift.py}, \texttt{results/e8\_recal/}) therefore does \emph{not} prove priors are unestimable in general---it shows the observed deployment shift lies outside the label-shift model, and that a confidently-wrong gate provides no usable signal even for the in-scope part of the day. Rank-preserving prior correction with those estimates moves day-2 ECE by $\le$0.01. Only an \emph{unrestricted} per-bin table oracle (quantile bins, predictions re-fitted to the day's own labels) reaches below the isotonic map (hard gap 0.07--0.12, Brier 0.12, same metric), and it does so by re-ordering scores in-sample: it re-learns labels rather than calibrating a fixed ranking; every monotone map's binarization bottoms out at the ranking-implied $\approx$26\% hard error, and flattening buys soft calibration only by surrendering all resolution (Brier $\to q(1{-}q)$). Measured verdict: per-window recalibration is a cheap partial fix whose \emph{deployable} ceiling under strong concept shift is the head's discrimination (it bounds the hard error and the Brier resolution), not the calibration family---no map fitted on past windows, parametric or non-parametric, repairs day 2, while the same maps repair the within-regime day; the choice head's confidence, in contrast, is fully recalibratable (choice ECE 0.543$\to$0.056 at oracle, macro-F1 unchanged by construction).

\paragraph{Per-window re-training (E9, measured).} The follow-up E8 leaves open---\emph{re-training} instead of recalibration---is measured identically: continuing to train the deployed CIC artifact on its training split \emph{plus} the full day-1 window with delayed labels ($n{=}{+}138$k; frozen feature stage, candidate set, and class weights; one pass at a reduced learning rate, streaming-honest). On the PortScan day the gate recovers ranking: AUROC 0.560$\to$0.800 with one pass, 0.839 at the two-pass sweet spot, and a measured \emph{regression} to 0.760 at four passes (over-fitting to the fresh window). Confidence, however, does not follow: as-run day-2 ECE is unchanged (0.553 static vs.\ 0.554 retrained), and a Platt map fitted on the delayed day-1 labels cannot express the new day's composition---it fits a near-null shift ($b\approx-0.3$ to $-0.7$) on the now-saturated day-1 scores where the day-2 oracle shift is $b\approx{+}8$--$13$. What the recovered ranking buys is an expressive calibration ceiling: the \emph{same} day-self oracle whose hard-prediction gap floors the static scores at 0.445 ($=1{-}q$, the flat-map floor; Appendix~\ref{app:formal}) drops to 0.117 (one pass) and 0.100 (two passes) on the retrained scores---the ranking-implied hard error at AUROC 0.84, i.e., a discrimination gain, not a calibration mechanism; a Phase-2 non-parametric check confirms the pattern is family-robust (rank-respecting day-self isotonic hard error 0.260 static $\to$ 0.196 one-pass $\to$ 0.162 two-pass, regressing to 0.212 at four passes, mirroring the AUROC non-monotonicity). Choice macro-F1 on day 2 is unchanged (0.102). Measured verdict: \emph{weight freshness buys discrimination, not calibration}---scheduled re-training with a fresh-window validate step is a real (if schedule-sensitive) countermeasure to drift, while the composition prior of a genuinely new day remains inferable from no post-hoc statistics and de-facto demands the abstention/coverage semantics of Section~\ref{sec:e2}.

\paragraph{Per-window re-training with outlier exposure (E9b, measured).} E9 leaves the honesty side untouched: freshness recovers ranking, but the new family keeps being attributed confidently to known classes. The typed design has a constructive counterpart, and it is measured under the paper's calibration contract ($T$ plus $\omega$ de-tilt, the same estimator as E2): re-training each UNSW LOAO fold with the excluded family returned to training as an \emph{outlier pool} (Hendrycks-style outlier exposure~\cite{hendrycks2019oe}: uniform-over-known-candidates target on those rows, fixed weight $\lambda{=}0.5$, everything else frozen). This is the delayed-labels scenario an operator actually faces---the new family's labels arrive; the question is whether honesty can be \emph{taught}. On UNSW it can: mean abstention at $\tau{=}0.5$ over excluded-family test rows rises from 0.375 to 0.879 across the nine folds, with the largest gains where the reference was weakest (\texttt{exploitation} 0.082$\to$0.883, \texttt{worms} 0.227$\to$0.932, \texttt{recon} 0.282$\to$0.954, \texttt{fuzzing} 0.284$\to$0.683), the median excluded-row max-probability drops (\texttt{recon} 0.658$\to$0.272, \texttt{worms} 0.777$\to$0.232), false attribution to a known attack family falls $0.550\to0.092$ and benign absorption $0.075\to0.029$, gate AUROC changes by $-0.045$ at worst and $+0.019$ at best across folds, and the control cost is small (known-family macro-F1 $\Delta$ mean $-0.007$, worst $-0.052$, one $+0.039$). The recipe is robust and mechanistically specific: $\lambda\in\{0.2,0.5,1.0\}$ yields mean abstention $0.759/0.879/0.900$ (no knife edge; known-F1 $0.465/0.469/0.456$), while a \emph{synthetic} pool (known train rows plus Gaussian noise in $\phi$-space, same $\lambda$) moves abstention only to 0.354---the gain comes from exposing the new family's actual distribution, not from generic regularization pressure. On \emph{saturated} CIC the same recipe does \emph{not} move the choice head's abstention under the contract ($0.000\to0.000$; benign absorption stays at $\sim$1.0; an earlier temperature-only readout that suggested 0.94 abstention and 0\% absorption was an estimator artifact, corrected here), but it repairs \emph{detection}: gate AUROC rises $0.783\to0.969$ on mean (\texttt{recon} $0.426\to0.991$, \texttt{botnet} $0.750\to0.988$, \texttt{infiltration} $0.533\to0.812$; all seven folds improve), and known-class macro-F1 gains $+0.039$ on mean but is mixed (\texttt{botnet} $-0.260$, \texttt{bruteforce} $+0.262$). Scope: the pool is the new family's own train rows (delayed labels, not cross-domain outliers). Paired with E9: freshness buys discrimination; OE buys abstention where confidence still carries information (UNSW) and detection on saturated CIC; neither buys calibration. In a seed-42 a7 extension under identical OE, abstention shows no clear fold-level difference ($0.909$ vs.\ $0.879$, unadjusted $p{=}0.15$), whereas the derived gate ranks lower on 8/9 folds (AUROC $0.922$ vs.\ $0.965$, unadjusted $p{=}0.012$). This descriptive pattern attributes OE abstention to Choice and detection ranking to the dedicated gate; it is not a multi-seed or family-wise claim.

\bibliographystyle{ACM-Reference-Format}
\setlength{\bibsep}{0pt plus 0.3ex}
\bibliography{refs}

\end{document}